# Quantitative magnetic imaging at the nanometer scale by Ballistic Electron Magnetic Microscopy

M. Hervé, S. Tricot, S. Guézo, G. Delhaye, B. Lépine, P. Schieffer and P. Turban[a]

*Département Matériaux et Nanosciences, Institut de Physique de Rennes, UMR 6251, CNRS-Université de Rennes 1, Campus de Beaulieu, Bât 11E, 35042 Rennes cedex, France*

We demonstrate quantitative ballistic electron magnetic microscopy (BEMM) imaging of simple model Fe(001) nanostructures. We use in situ nanostencil shadow mask resistless patterning combined with molecular beam epitaxy deposition to prepare under ultra-high vacuum conditions nanostructured epitaxial Fe/Au/Fe/GaAs(001) spin-valves. In this epitaxial system, the magnetization of the bottom Fe/GaAs(001) electrode is parallel to the [110] direction, defining accurately the analysis direction for the BEMM experiments. The large hot-electron magnetoresistance of the Fe/Au/Fe/GaAs(001) epitaxial spin-valve allows us to image various stable magnetic configurations on the as-grown Fe(001) microstructures with a high sensitivity, even for small misalignments of both magnetic electrodes. The angular dependence of the hot-electron magnetocurrent is used to convert magnetization maps calculated by micromagnetic simulations into simulated BEMM images. The calculated BEMM images and magnetization rotation profiles show quantitative agreement with experiments and allow us to investigate the magnetic phase diagram of these model Fe(001) microstructures. Finally, magnetic domain reversals are observed under high current density pulses. This opens the way for further BEMM investigations of current-induced magnetization dynamics.

a) Author to whom correspondence should be addressed. Electronic mail: pascal.turban@univ-rennes1.fr

I. INTRODUCTION

Understanding the magnetic properties of low-dimensional systems is one major issue for the future development of magneto-electronics devices with high areal density. At low scale, the presence of magnetic domains plays a crucial role on the magnetic response of small ferromagnetic elements[1]. Recent development of magnetic microscopy tools for the analysis of bulk materials as well as thin films, surfaces, interfaces and nanostructures[2] has allowed a direct comparison of micro- and nano-magnetic models with experiments, down to the nano-scale. Among magnetic microscopies presenting high lateral resolution, the recently proposed Ballistic Electron Magnetic Microscopy (BEMM)[3,4] appears promising due to some of its specific features. In BEMM, a scanning tunneling microscope (STM) non-magnetic tip is used to locally inject a hot electron current in a spin-valve grown on a semiconducting substrate. The ballistic current $I_C$ collected in the semiconductor is modulated by the hot-electron giant magneto-resistance (GMR) effect, allowing magnetic imaging of the spin-valve domain structure. BEMM is thus sensitive to the volume magnetism of thin films and nanostructures, and can be used to image the magnetic domain structure of buried ultrathin elements integrated in realistic magneto-resistive devices. Although a measurement of the ultimate BEMM magnetic resolution is still missing, an upper bound of 28nm was reported on p-doped samples[5], while a lateral resolution below 1nm was reported in ballistic electron emission microscopy experiments on non-magnetic samples[6]. Up to now, BEMM was quasi exclusively used to investigate magnetic domain structure or magnetization reversal process in planar spin-valves only. Most of these studies were limited to a qualitative analysis of the parallel and anti-parallel magnetic domain evolution under applied field, and only recently BEMM quantitative analysis of magnetic domain walls profiles was reported[7,8]. Despite the BEMM high lateral resolution, only one study was dedicated to the investigation of FeNi patterned magnetic nanostructures[4]. This is most likely due to the fact that classical UV or electron lithography is hardly compatible with the preservation of a post-process surface cleanliness allowing stable STM/BEMM imaging conditions. In this paper, we report on BEMM investigation of ultrathin epitaxial Fe(001) microstructures integrated in epitaxial Fe/Au/Fe/GaAs(001) spin-valves. We use in situ shadow mask deposition to pattern elongated magnetic elements in the top Fe(001) spin-valve electrode. Competition between shape anisotropy, magneto-crystalline anisotropy and exchange energy terms results in the formation of various magnetic domain structures on the as-grown sample. These various magnetic configurations are reproduced by micromagnetic calculations. Furthermore, we demonstrate that the calculated magnetization maps can be converted into simulated BEMM images. An excellent agreement is obtained between simulated and measured BEMM images demonstrating the BEMM ability for quantitative magnetic imaging of micro/nanostructures. Finally, we report observation of magnetic domain wall motion under high injected current pulses. This suggests that BEMM can also be used as a powerful experimental tool for the investigation of current-induced magnetization dynamics in micro/nanostructures.

## II. EXPERIMENTAL DETAILS

Samples are prepared by molecular beam epitaxy (MBE). A 1.5µm thick Si n-doped ($4 \times 10^{16}$cm$^{-3}$) GaAs buffer layer is first grown in a independent MBE chamber on a n$^+$-GaAs(001) substrate and protected by a 5µm thick amorphous As capping layer to allow transfer under ambient atmosphere in the BEMM setup. The ultra-high vacuum (UHV) BEMM setup[9], with a base pressure better than $1 \times 10^{-10}$mbar, consists of a metal MBE deposition chamber connected to the STM/BEMM microscope. Thermal desorption of the As protecting layer is first done at 760K, in front of a cryopanel cooled with liquid nitrogen, leading to the formation of a clean As(2×4)-reconstructed

GaAs(001). After sample cooling down to room temperature, MBE deposition of the Fe(1nm)/Au(6nm)/Fe(1.2nm)/GaAs(001) spin-valve on the GaAs surface is performed by using two Knudsen cells. Two levels of shadow-mask microstructuration are used during the spin-valve deposition process. The bottom Fe electrode and the non-magnetic Au spacing layer are first deposited through a shadow mask to form 400μm diameter metallic dots on the GaAs(001) substrate. This allows us to obtain highly resistive Au/Fe/GaAs(001) Schottky contacts (R>10MΩ at room temperature) for low-noise BEMM measurement. The top Fe electrode of the spin-valve is finally patterned in an array of 1700*550nm$^2$ capsule-shaped elements via the nanostencil technique[10,11]. The mask, a 900 nm thick silicon nitride membrane structured by focused ion beam, is gently brought into direct contact with the GaAs substrate, and deposition is performed with normal incidence to minimize blurring of the structure edges. In the following, all crystalline directions will refer to crystal directions of the GaAs(001) substrate. The long axis of the Fe microstructure was aligned along [1-10]. Electrochemically etched W STM tips are cleaned *in situ* by thermal heating before the BEMM experiments. The hot-electron current $I_C$ is collected at the back of the GaAs substrate using an indium ohmic contact. A gold wire mounted on a translation motion and a gimbal tilter is used to ground selectively the surface of the 400μm diameter isolated metallic dots[12] in the STM/BEMM microscope head. All STM/BEMM experiments are performed at room-temperature in the constant-current mode of operation, without applying any magnetic field on the as grown spin-valve. Micromagnetic simulations are performed using the freely available OOMMF package[13] (Object Oriented MicroMagnetic Framework). The Gilbert gyromagnetic ratio is set to 2.2x10$^5$ m/A/s and the damping constant is kept large (α=0.5) to ensure fast simulation convergence. The magnetic volume is discretized using a 2.5x2.5x1 nm$^3$ cell. The Fe stiffness constant is set to 1.13x10$^{-11}$ J/m[14] and the saturation magnetization to 1720 kA/m. A slight uniaxial anisotropy energy contribution (with a [110] easy axis) is introduced as well as a biaxial anisotropy energy term ([100] and [010] easy axes) with respective magneto-crystalline anisotropy constants $K_u$=2.73 kJ/m$^3$ and $K_1$=8kJ/m$^3$, as measured by magneto-optical Kerr effect on identical unpatterned Fe film. The resulting magnetic anisotropy defines two easy axes, respectively 30° and 150° away from the [110] direction for an unpatterned planar top Fe electrode. No magnetic coupling was considered vertically between both Fe(001) electrodes of the spin-valve due to the large Au spacer thickness nor laterally between neighboring dots.

## III. RESULTS AND DISCUSSION

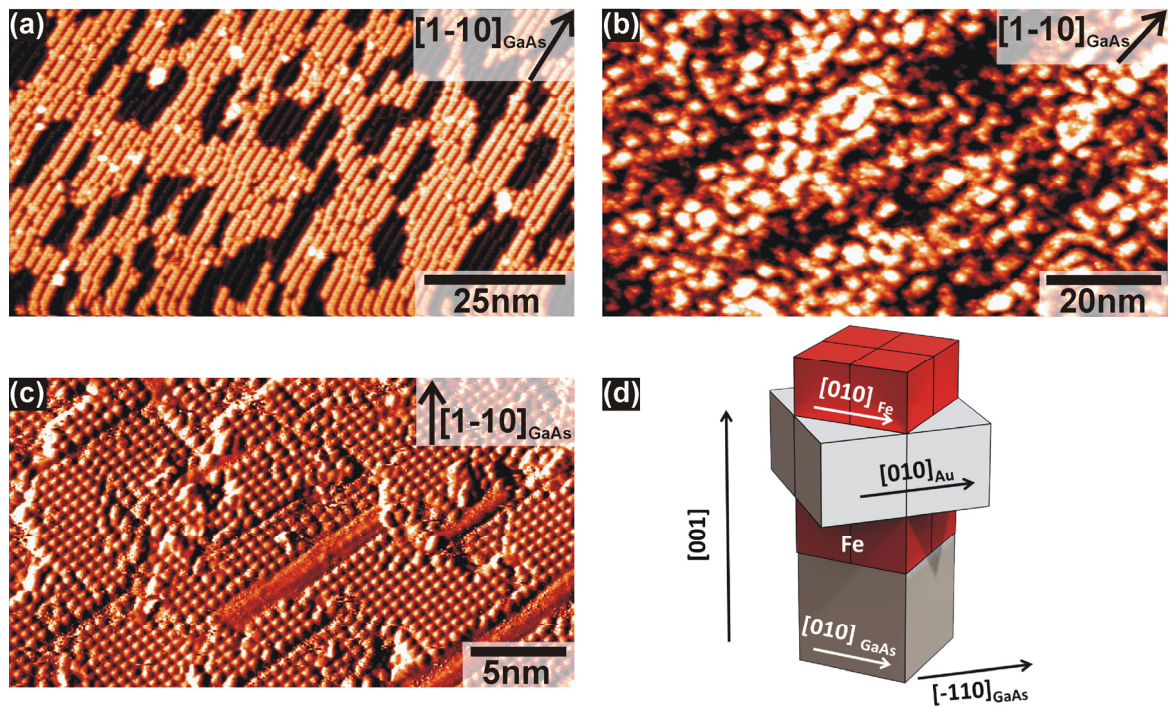

FIG. 1. Surface morphology at each step of the spin-valve deposition : (a) 100*55nm$^2$ STM image of the GaAs(001) surface after desorption of the protective amorphous As cap. (b) 100*55nm$^2$ STM image after deposition of the 1.2nm thick Fe(001) bottom electrode. (c) 30*16nm$^2$ STM image of the 6nm thick Au(001) spacing layer. (d) epitaxial relationship of the Fe/Au/Fe/GaAs(001) stack.

Fig. 1 displays STM images recorded at each step of an unpatterned epitaxial Fe(1nm)/Au(6nm)/Fe(1.2nm)/GaAs(001) spin-valve deposited at room temperature. After the As capping layer thermal desorption, the GaAs(001) initial surface is As-rich, and presents a disordered As(2*4) surface reconstruction, with As dimer lines running along the [1-10] GaAs crystal direction (Fig. 1(a)). The obtained GaAs surface is atomically flat over large areas, with a typical atomic terraces size of 400nm. After deposition of the 1.2nm thick Fe bottom electrode, a surface granular structure is observed (Fig. 1(b)). The surface is made of few nanometer wide atomically flat Fe 3D islands which appear slightly elongated along the [1-10] crystal direction. Despite the small size of these Fe terraces, the resulting peak-to-peak roughness is limited to 0.4nm over 400*400nm$^2$ areas. Deposition of the 6nm thick Au spacer leads to a significant smoothing of the sample surface (Fig. 1(c)), with Au atomic terraces expanding over 30nm and presenting a (2*2) surface reconstruction. This surface reconstruction is due to the segregation of roughly half a monolayer of As as confirmed by X-ray photoelectron spectroscopy experiments. As is well known to segregate from the GaAs surface during the Fe layer deposition[15] and is further floating on the surface during the Au spacer deposition. Finally, the top 1nm thick Fe electrode presents a surface morphology similar to the first electrode. The obtained epitaxial relationship in the stack is Fe[100]//Au[110]//Fe[100]//GaAs[100] (Fig. 1(d)).

Fig. 2(a) displays a typical 1.4*1.4μm$^2$ large scale STM image of the final spin-valve, with the top Fe electrode patterned by the nanostencil technique. Atomic steps of the GaAs(001) substrate are observed after the spin-valve deposition. One Fe(001) dot (top electrode) is observed with its long axis parallel to the [1-10] direction of the GaAs substrate. The measured height of the Fe dot is 1.0±0.1nm in agreement with its targeted thickness. A line profile over the Fe(001) dot edges shows an edge broadening with a width of 50nm, typical of the nanostencil technique. Fig. 2(b) presents the BEMM image recorded simultaneously with the STM image. This image corresponds to a (80 pixels*80 pixels) grid of the hot electron current $I_C$ recorded over the (400 pixels*400 pixels) STM image. Tunnel current was set to 20nA and hot-electron energy to 2.1eV in order to improve the signal-to-noise ratio. The magnetization of the bottom Fe electrode is parallel to the [110] direction of the substrate due to the strong uniaxial magnetic anisotropy at the Fe/GaAs(001) interface[16], defining the analyzer direction for the BEMM experiments.

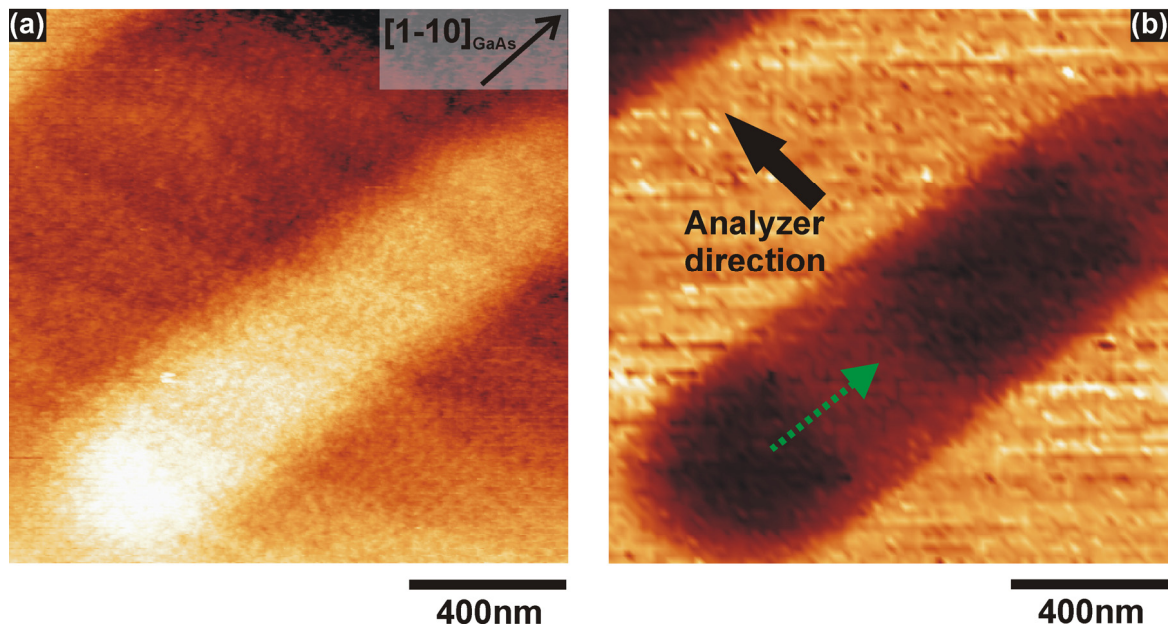

FIG. 2. (a) Left : 1.4*1.4μm² STM image of the Fe(1nm)/Au(6nm)/Fe(1.2nm)/GaAs(001) spin-valve. The 1700*550nm² capsule-shaped top Fe(001) electrode has its long axis parallel to the [1-10] direction of the GaAs substrate. (b) Right: corresponding BEMM image (color scale: 6 to 157pA). Tunneling current was set to 20nA and electron energy to 2.1eV. Black arrow indicates the orientation of the Fe bottom electrode (BEMM analyzer), parallel to [110].

On the BEMM image, a sequence of dark (low ballistic current) and bright (high ballistic current) areas are observed along the top Fe dot long axis, without any correlation with the surface morphology. This contrast is from magnetic origin and attributed to the presence of four magnetic domains in the microstructure. The corresponding magneto-current (MC) at 2.1eV between dark and bright BEMM areas (defined as $\frac{I_C(\text{bright}) - I_C(\text{dark})}{I_C(\text{dark})}$) is of 115%. This value is smaller than the MC amplitude measured on an identical unpatterned planar spin-valve ($MC_{180°}$=260% of relative change of the BEMM current at 2.1eV, between a strictly parallel and strictly anti-parallel magnetic configuration of both Fe electrodes), suggesting that the magnetization rotation angle between dark and bright regions of Fig. 2(b) is smaller than 180°. A halo of higher BEMM current is also observed all around the Fe(001) dot. The width of this brighter area is of typically 80nm and is thus related to the Fe wedge at the borders of the microstructure. This wedge causes an increase of the hot-electron intensity due to the reduced Fe thickness crossed by the hot-electron beam. The magneto-crystalline anisotropy of the Fe(001) layer, and thus the magnetization orientation close to the borders, is also modified when the film thickness decreases from 1 to 0 nm. A part of the halo can thereby also have a magnetic origin, but a clear description of this specific part of the BEMM image is beyond the scope of this paper. In the following, abrupt edges of the microstructure are considered for micromagnetic simulations.

Fig. 3(a) presents the calculated stable micromagnetic configuration obtained for the Fe(001) capsule. Before calculation, the dot was arbitrarily initialized in a 4 magnetic domains configuration whith the magnetization lying along the easy axis of the unpatterned top electrode. The simulation stabilizes 94° Néel walls and the magnetization in the dot is 43° (respectively 137°) away from the [110] analyzer direction for the bright (respectively dark) domains. This is a consequence of the introduction of supplementary shape anisotropy by the patterning process: both easy axes of the system rotate by 13° in direction of the long axis of the microstructure. For a quantitative comparison of the micromagnetic simulation with BEMM experiments, we used the magnetization map of Fig. 3(a) as an input to simulate a BEMM image. The hot-electron current depends on the angle θ between the local magnetization of the Fe dot and analyzer direction with a $\sin^2\left(\frac{\theta}{2}\right)$ law[17].

We could thus compute the BEMM current at each cell of the OOMMF simulation by considering the angle θ between local magnetization direction in the Fe dot and the reference [110] direction of the Fe bottom electrode. The obtained simulated BEMM image (Fig. 3(b)) is nicely matching the experimental BEMM image of Fig. 2(b). An experimental BEMM current profile $I_C(x,y)$ across the 94° Néel wall was also extracted from Fig. 2(b) (dotted green arrow), and converted into an angular profile θ(x,y). A quantitative agreement (Fig. 3(c)) is obtained with the micromagnetic simulation, without using any free parameter nor introducing any experimental broadening. The measured width of the 94° domain wall is of 130nm.

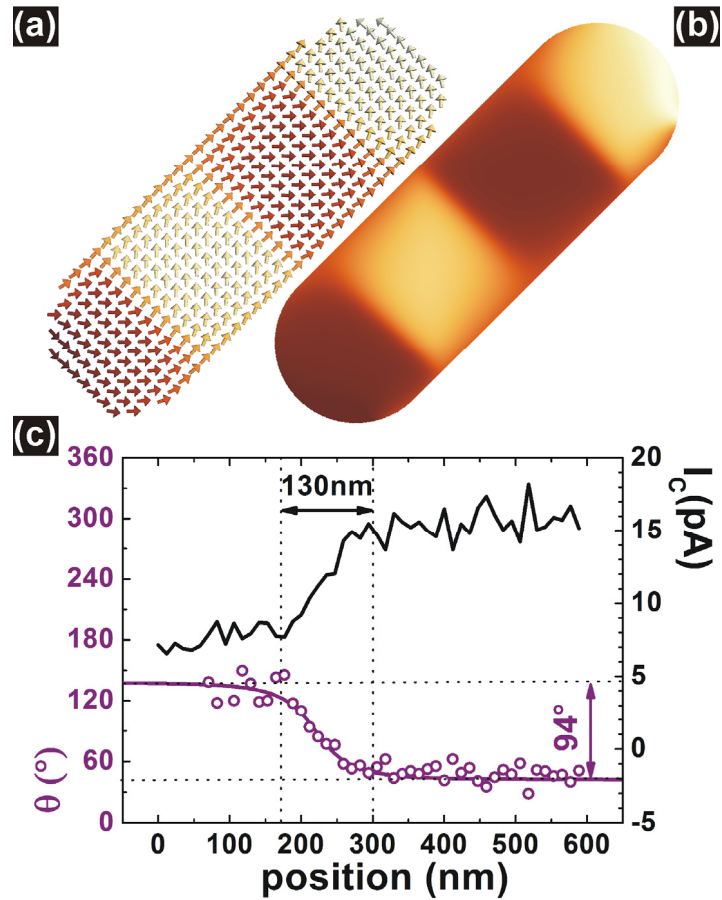

FIG. 3. Quantitative analysis of the experimental BEMM image from Fig. 2(b): (a) micromagnetic calculation of the magnetization map in the Fe(001) dot. (b) Corresponding calculated BEMM

image. (c) Experiment/simulation comparison: plot of a measured BEMM current profile across a magnetic domain wall (black continuous line); this profile was extracted along the green dotted arrow of Fig. 2(b) and converted in an experimental magnetization angle profile (purple empty circles). Continuous purple line corresponds to the simulated magnetization angle profile obtained by micromagnetic calculation. Magnetization rotates by 94° across the 130nm wide domain wall.

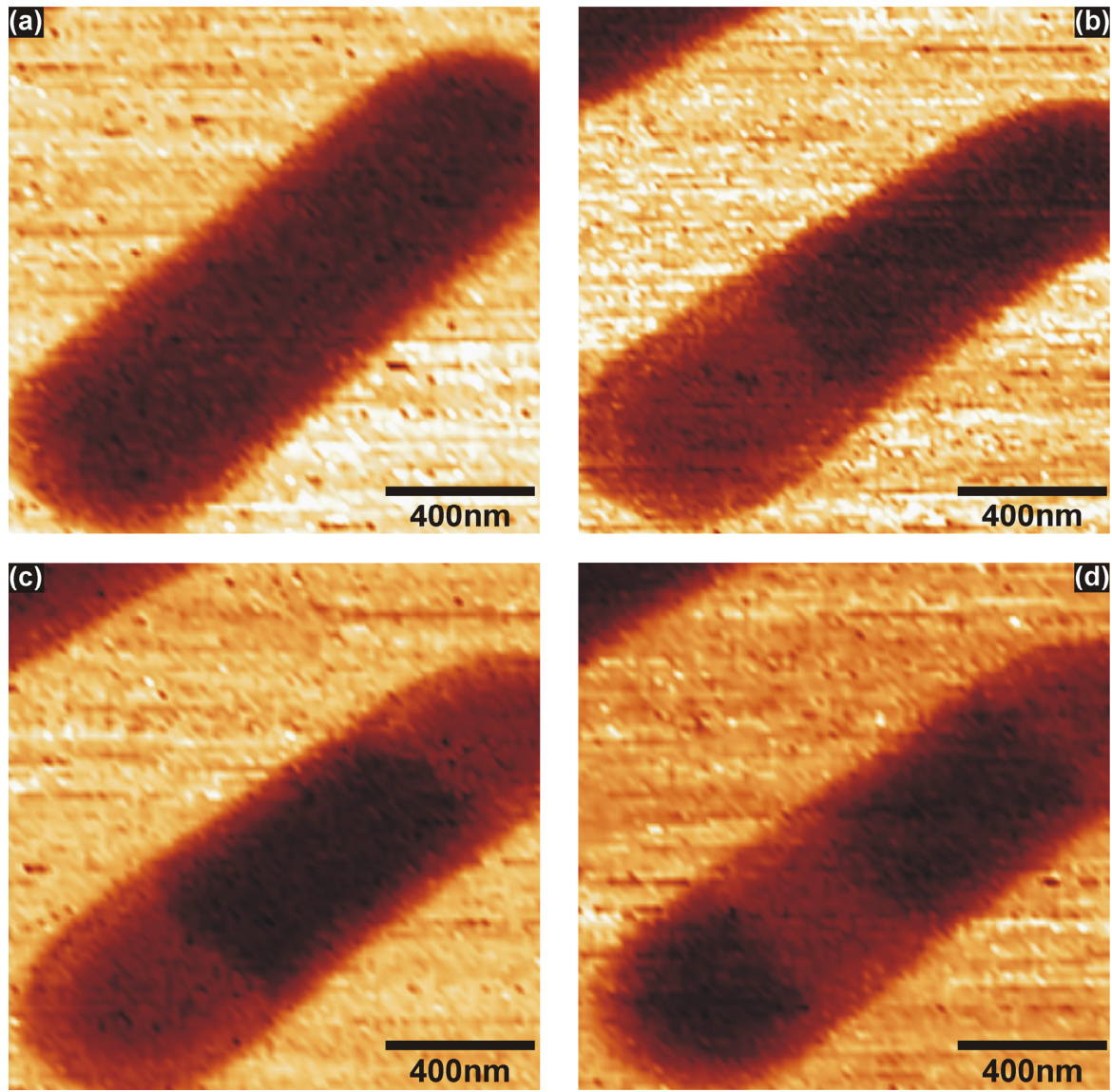

FIG. 4. 1.4*1.4μm² BEMM images of the 4 only magnetic configurations observed after sampling 30 as-grown Fe(001) dots, presenting 1 (a), 2 (b), 3 (c) or 4 (d) magnetic domains.

30 other identical Fe(001) dots were probed over the sample surface. Only four kind of magnetic configurations were observed by BEMM, presenting 1, 2, 3 or 4 magnetic domains, as reported on Fig. 4(a)-4(d). All the bright (respectively dark) domains present the same $I_C$ values as the dark/bright areas of Fig. 2(b), and correspond once more to magnetization orientation 43° (respectively 137°) away from the [110] analyzer direction. For all observed configurations, the number and position of magnetic domain walls in the microstructure do not show any obvious correlation with the underlying GaAs substrate atomic steps. The domain walls also do not present any important bowing. We thus don't have experimental evidence of domain wall pining by structural defects and consider that the observed magnetic configurations are among the lowest-energy configurations of the magnetic phase diagram. To check this point, we reproduced the observed configurations by micromagnetic calculations. The microstructure was arbitrarily initialized in a configuration with 1, 2, 3 or 4 magnetic domain with magnetization parallel to the easy axis of the Fe unpatterned top electrode. Except for the single domain case, calculations stabilized multi-domains structures separated by 94° domain walls. The final positions of the domain walls in the stable calculated magnetic configurations are in good agreement with the BEMM images of Fig. 4(a)-4(d). The corresponding total magnetic energies are plotted in Fig. 5. The 4 configurations which are experimentally observed by BEMM correspond to the lowest calculated magnetic energies and the absolute minimum magnetic energy is obtained for the two domains structure. The energy difference between the 1, 2 and 3-domains states is weak (0.03kJ/m$^3$ corresponding to 158meV for one dot) and accounts for the majority presence of these configurations on the as-grown sample. The 4-domains configuration was observed only once over 30 sampled dots, which is coherent with its significantly higher energy (551meV higher than the 2-domains state).

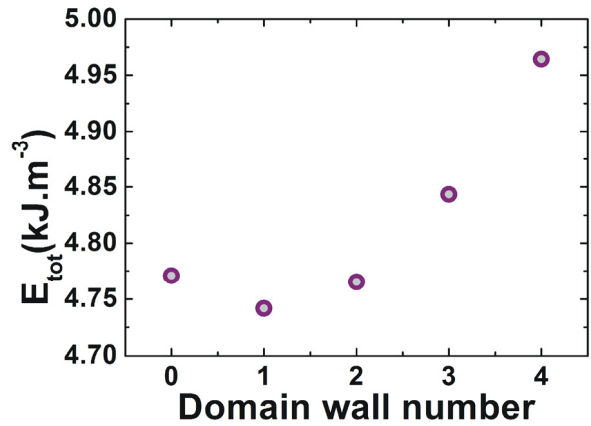

FIG. 5. Plot of the total magnetic energy calculated by micromagnetic simulations for magnetic states of the Fe(001) dot with 0, 1, 2, 3 or 4 magnetic domain walls. The 4 lowest energy states correspond to the magnetic configurations experimentally observed in Fig. 4.

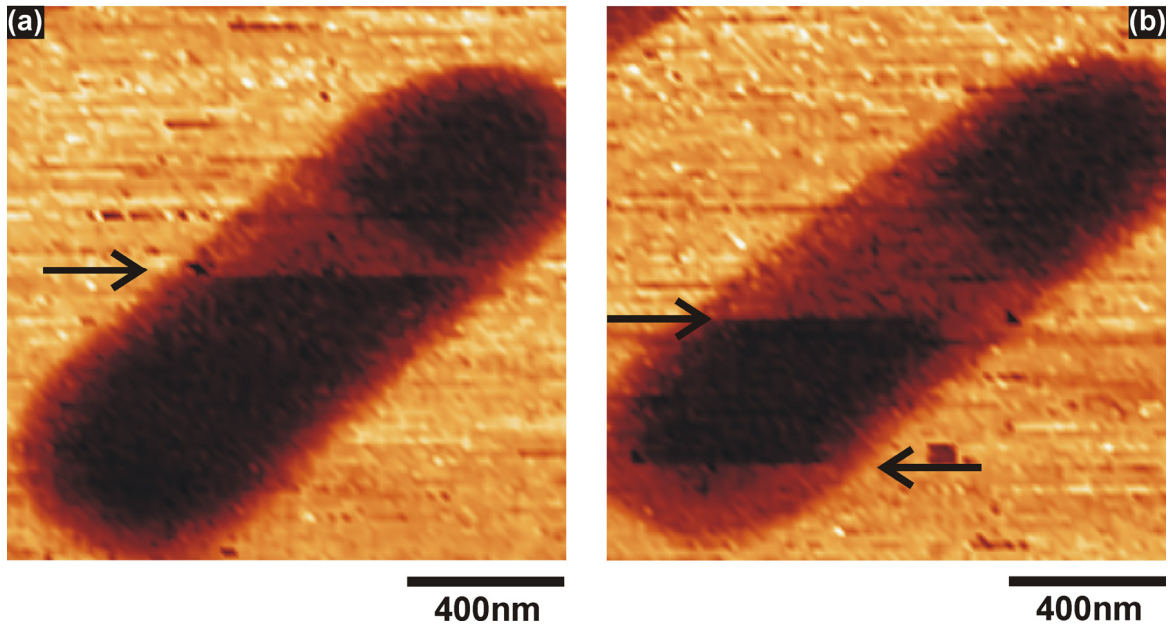

FIG. 6. 1.4*1.4µm$^2$ BEMM images of Fe(001) dots. Scan direction is from left to right and bottom to top. STM tip instabilities were observed while scanning and generated high current density pulses (typically in the 100-200nA range). These events are marked by black arrows and cause magnetization switching between the two easy axis of the Fe(001) dot as evidenced by BEMM contrast inversion.

Finally, some unexpected events were observed during the acquisition of the BEMM images. Due to the high tunnelling current values ($I_T$=20nA) and tunnel bias used in the BEMM experiments, STM tip instabilities are common while scanning (mostly due to material exchange between tip and surface). These instabilities are accompanied by local injection of intense tunnel current pulses (up to 200nA) in the spin valve. We noticed that when these current pulses were located inside the Fe dots, a sudden change in the BEMM signal (marked by arrows on Fig. 6(a)-6(b)) was observed inside the magnetic microscturcture, without any signal change outside the dot. We checked that similar current pulses generated outside the Fe dots did not lead to any similar contrast changes. This contrast change is thus from magnetic origin and related to current-induced magnetic domain reversal. Various driving-forces can lead to current induced magnetic switching. In our experiments, the Oersted field locally generated during current injection can be estimated typically of the order of few ten of µT, i.e. of the same order of magnitude than earth magnetic field. The influence of this Oersted field alone is thus questionable. Local Joule heating of the sample can promote domain wall motion, when supplementary thermal energy allows detrapping of the domain wall from structural defects. Recent spin-polarized experiments[18] demonstrated that the Joule heating of a nanometric Fe/W(110) monoatomic island induced by electron tunnelling at 40K led to a typical effective temperature increase of 17K/µW for a typical 4*4nm$^2$ island. This would correspond in our experiments to an estimated upper bound for the local temperature increase under the STM tip of 1K considering a 10*10nm$^2$ heat dissipation area. This temperature increase is low compared to the thermal energy at room temperature. Since we don't observe any thermal fluctuation of the magnetic configuration with a stable tip, the influence of the local Joule heating alone can be excluded. Finally, spin-transfer torque effect can promote magnetization rotation induced by current pulses in such spin-valve structure[19]. The current pulse during the tip instability can potentially achieve a current density up to $10^{11}$ A.m$^{-2}$ (considering a 100nA pulse over a 1*1nm$^2$ injection area) which is sufficient for spin-transfer torque observation. However, when the electron beam is flowing from the free magnetic layer (microstructure) to the fixed magnetic layer (bottom Fe electrode) the spin transfer torque effect should promote a switching towards the anti-parallel magnetic state of the spin-valve. As can be seen on Fig. 6(b), we observe both kinds of magnetic contrast inversions, either from the parallel to the anti-parallel state or vice versa. From this very preliminary analysis, we can thus not clearly attribute a unique and simple origin to the observed magnetization switching phenomenon. A more complex combination of thermal, Oersted field and spin-transfer torque effects may account for our observations, but further detailed experimental investigations combined with micromagnetic simulations is necessary to conclude on this point. However, these primary observations demonstrate the potential interest of BEMM for the investigation of current-induced magnetization dynamics.

## IV. CONCLUSIONS

In conclusion, we observe by ballistic electron magnetic microscopy the magnetic domain structure of epitaxial Fe(001) sub-micron elements integrated in epitaxial Fe/Au/Fe/GaAs(001) spin valve. We take advantage of the high uniaxial magnetic anisotropy of the bottom Fe/GaAs(001) electrode to quantitatively analyze the magnetization orientation of the unknown Fe magnetic microstructure with respect to the [110] direction. The large magneto-contrast between parallel and anti-parallel states of the spin-valve gives access to a large dynamical range of BEMM current for a 360° rotation of the magnetization. Angular dependence of the magneto-contrast is used to convert micromagnetic simulations into simulated BEMM images in quantitative agreement with experiments. A 13° rotation of the magnetic easy axes caused by shape anisotropy in the Fe(001) dots is determined by BEMM. The various magnetic configurations imaged by BEMM on the as-grown sample are finally matching the lowest energy states of the magnetic phase diagram for the system. A further step of this BEMM study is to investigate in situ the magnetization reversal mechanisms of the microstructure, either by applying an in-plane magnetic field on the sample, or by using the STM tip to locally generate in a controlled manner high current density pulses. This work is in progress.

## ACKNOWLEDGMENTS

This work was financially supported by Région Bretagne and Rennes Métropole. Technical support by Arnaud Le Pottier during the BEMM setup development is gratefully acknowledged.